\documentclass[12pt]{article}%
\usepackage{citesort}
\usepackage{amsmath}
\usepackage{graphicx}%
\usepackage{amsfonts}%
\usepackage{amssymb}
\def\ba{\begin{eqnarray}}
\def\ea{\end{eqnarray}}

\begin{document}

\title{Constituent quark and diquark properties from small angle proton-proton
elastic scattering at high energies}
\author{A.Bialas and A.Bzdak\thanks{Fellow of the Polish Science Foundation (FNP)
scholarship for the year 2006.}\\M.Smoluchowski Institute of Physics \\Jagellonian University, Cracow\footnote{Address: Reymonta 4, 30-059 Krakow,
Poland; e-mail: bialas@th.if.uj.edu.pl}}
\maketitle

\begin{abstract}
Small momentum transfer elastic proton-proton cross-section at high energies
is calculated assuming the nucleon composed of two constituents - a quark and
a diquark. A comparison to data (described very well up to $-t \approx2$
GeV$^{2}$/c) allows to determine some properties of the constituents. While
quark turns out fairly small, the diquark appears to be rather large,
comparable to the size of the proton.

\end{abstract}

\section{Introduction}

The quark structure of hadrons at low momentum transfers can manifest itself
in many physical phenomena. It is very important for their static properties
as, e.g., magnetic moments and mass relations which are reasonably well
described by the constituent quark model \cite{mm}. It can be used for
description of the elastic amplitudes at low momentum transfers and spin
effects in two-body processes \cite{el}. Also, as it was suggested long time
ago \cite{bcf}, and rediscovered recently \cite{reszta}, the quark structure
of nucleon may be crucial in analysis of particle production from nuclear targets.

In most applications of the quark model to low-momentum transfer phenomena
only single $qq$ interactions are considered. In this case the possible
correlations between constituents are unimportant. When multiple scattering is
taken into account \cite{bj}, however, the effects of correlations cannot be
neglected.\footnote{An indication that they may indeed be necessary to account
for the precise data on elastic $pp$ scattering can be inferred from
\cite{pp-3q}.}

One possibility to introduce correlations between the constituent quarks
inside the nucleon is to combine two of the quarks into one object, a diquark
\cite{Wilczek}. This is the possibility we explore in the present
investigation: we discuss the elastic nucleon-nucleon scattering, assuming
that the nucleon is composed of two constituents - a quark and a diquark. Our
main goal is to determine the properties of these two constituents by
comparing the results of calculations with data. They are needed for the
analysis of RHIC data on particle production from nuclei \cite{phobos} in the
''wounded'' \cite{bbc} constituent model \cite{bb}.

In its most general formulation, the quark-diquark model assumes that the
nucleon consists of two constituents (quark and diquark), acting
independently. This assumption must be, surely, supplemented by a more
detailed description of the specific properties of constituents and of the
distribution of constituents inside the nucleon. Thus the model is rather
flexible and one should not be surprised that it can be adjusted to describe
correctly the data. The real interest is that, when confronted with data, the
model can provide information on the details of nucleon structure at small
momentum transfers.

This is exactly the logic behind the present investigation. We first verify if
the model can account for the data on low momentum transfer elastic
proton-proton scattering at high energies. We found that this is indeed the
case and that one obtains a really excellent description of data. This allows
to discuss the main point of this paper: the properties of the two
constituents and their distribution inside the nucleon.

Two formulations, differing by the treatment of the diquark structure, were
considered. One treats the diquark as a single object, the second one as an
object composed of two constituent quarks. Both gave rather similar results,
indicating that our conclusions are not sensitive to the details of the model.
The most spectacular outcome is that the diquark turns out to be rather large,
comparable to the size of the nucleon itself. Other conclusions are discussed
in the last section.

In the next section the general formulation of the model and its consequences
for high-energy small momentum transfer scattering are described. Two specific
examples of the implementation of these general ideas are presented (and the
corresponding results discussed) in Sections 3 and 4. In the last section our
conclusions are listed and commented.

\section{Low momentum transfer elastic scattering in the quark-diquark model}

\begin{figure}[h]
\begin{center}
\includegraphics[scale=1]{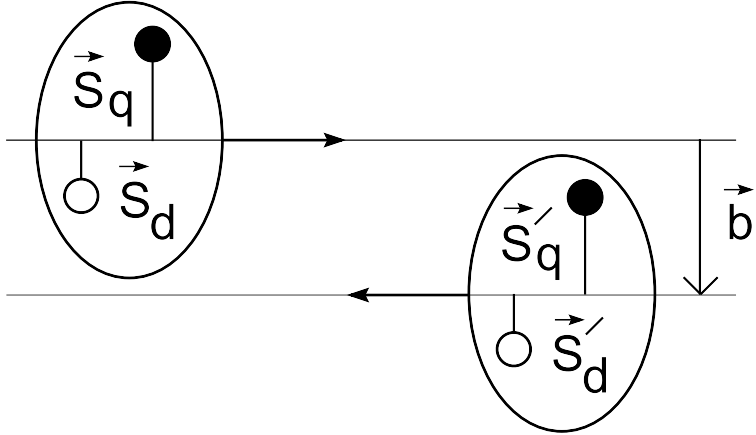}
\end{center}
\caption{Proton-proton scattering in the quark-diquark model.}%
\label{Fig_pp}%
\end{figure}

We follow the standard point of view that the imaginary part of the elastic
scattering amplitude, dominating at high energy, is generated by the
absorption of the incident particle wave, represented by the inelastic
(non-diffractive) collisions. The inelastic proton-proton cross-section at a
fixed impact parameter $b$, $\sigma(b)$, is calculated using the rules of the
probability calculus. One writes%
\begin{equation}
\sigma(b)=\int d^{2}s_{q}d^{2}s_{q}^{\prime}d^{2}s_{d}d^{2}s_{d}^{\prime
}D(s_{q},s_{d})D(s_{q}^{\prime},s_{d}^{\prime})\sigma(s_{q},s_{d}%
;s_{q}^{\prime},s_{d}^{\prime};b), \label{sigma}%
\end{equation}
where $D(s_{q},s_{d})$ denotes the distribution of quark and diquark inside
the nucleon, $s_{q}(s_{q}^{\prime})$, $s_{d}(s_{d}^{\prime})$ are transverse
positions of the quarks and diquarks in the two colliding nucleons, and
$\sigma(s_{q},s_{d};s_{q}^{\prime},s_{d}^{\prime};b)$ is the probability of
interaction at fixed impact parameter and the transverse positions of all
constituents taking part in the process. This configuration is illustrated in
Fig. \ref{Fig_pp}. Since the constituents act independently we have
\cite{cm,Glauber}:%
\begin{align}
1-\sigma(s_{q},s_{d};s_{q}^{\prime},s_{d}^{\prime};b)  &  =[1-\sigma
_{qq}(b+s_{q}^{\prime}-s_{q})][1-\sigma_{qd}(b+s_{d}^{\prime}-s_{q}%
)]\nonumber\\
&  [1-\sigma_{dq}(b+s_{q}^{\prime}-s_{d})][1-\sigma_{dd}(b+s_{d}^{\prime
}-s_{d})],
\end{align}
where $\sigma_{ab}(s)\equiv d^{2}\sigma_{ab}(s)/d^{2}s$ are inelastic
differential cross-sections of the constituents ($ab$ denotes $qq$, $qd$ or
$dd$).

For the distribution of the constituents inside the nucleon we take a Gaussian
with radius $R$:
\begin{equation}
D(s_{q},s_{d})=\frac{1+\lambda^{2}}{\pi R^{2}}e^{-(s_{q}^{2}+s_{d}^{2})/R^{2}%
}\delta^{2}(s_{d}+\lambda s_{q}). \label{D}%
\end{equation}
The second parameter, $\lambda$, has the physical meaning of the ratio of the
quark and diquark masses $\lambda=m_{q}/m_{d}$ (the delta function guarantees
that the center-of-mass of the system moves along the straight line). One
expects of course $1/2\leq\lambda\leq1$.

Our strategy is to adjust the parameters of the model by demanding that (i)
the total inelastic cross section (ii) the slope of the elastic cross section
at $t=0$ (iii) the position of the first diffractive minimum in elastic cross
section and (iv) the height of the second maximum in elastic scattering are in
agreement with data.

The unitarity condition implies for the elastic amplitude\footnote{Here and in
the following we are ignoring the real part of the amplitude.}
\begin{equation}
t_{el}(b)=1-\sqrt{1-\sigma(b)}. \label{unitarity}%
\end{equation}
The elastic amplitude in momentum transfer representation $T(\Delta)$ is a
Fourier transform of the amplitude in impact parameter space:%
\begin{equation}
T(\Delta)=\int t_{el}(b)e^{i\vec{\Delta}\cdot\vec{b}}d^{2}b=2\pi\int
t_{el}(b)J_{0}(\Delta b)bdb,
\end{equation}
where $J_{0}$ is the Bessel function.

With this normalization one can evaluate the relevant measurable quantities.
Total cross section:%
\begin{equation}
\sigma_{tot}=2T(0),
\end{equation}
elastic differential cross section ($t\simeq-|\Delta|^{2}$):%
\begin{equation}
\frac{d\sigma}{dt}=\frac{1}{4\pi}|T(\Delta)|^{2},
\end{equation}
slope of the elastic cross section (at $t=0$):%
\begin{equation}
B\equiv\frac{d}{dt}\left(  \ln\frac{d\sigma}{dt}\right)  |_{t=0}=\frac{\int
t_{el}(b)b^{3}db}{2\int t_{el}(b)bdb}. \label{B}%
\end{equation}

In the next sections we discuss two different choices for inelastic
differential cross-sections of the constituents $\sigma_{ab}(s)$.

\section{Diquark as a simple constituent}

As a first choice we parametrized $\sigma_{ab}(s)$ using simple Gaussian
forms:%
\begin{equation}
\sigma_{ab}(s)=A_{ab}e^{-s^{2}/R_{ab}^{2}}. \label{pab}%
\end{equation}
The radii $R_{ab}$ were constrained by the condition $R_{ab}^{2}=R_{a}%
^{2}+R_{b}^{2}$ where $R_{a}$ denotes the quark or diquark's radius (a natural
constraint for the Gaussians).

This means that the quark and the diquark are treated on the same footing,
their internal structures being described by one parameter, the radius
($R_{q}$ or $R_{d}$).

From (\ref{pab}) we deduce the total inelastic cross sections: $\sigma
_{ab}=\pi A_{ab}R_{ab}^{2}$. To reduce the number of parameters, we also
demand that the ratios of cross-sections satisfy the condition:%
\begin{equation}
\sigma_{qq}:\sigma_{qd}:\sigma_{dd}=1:2:4, \label{124}%
\end{equation}
expressing the idea that there are twice as many partons in the constituent
diquark than those in the constituent quark (shadowing neglected). This allows
to evaluate $A_{qd}$ and $A_{dd}$ in terms of $A_{qq}$:%
\begin{equation}
A_{qd}=A_{qq}\frac{4R_{q}^{2}}{R_{q}^{2}+R_{d}^{2}},\quad A_{dd}%
=A_{qq}\frac{4R_{q}^{2}}{R_{d}^{2}}. \label{A=A}%
\end{equation}

One sees that the model in this form contains $5$ parameters $R$, $\lambda$,
$R_{q}$, $R_{d}$ and $A_{qq}$ (we expect $A_{qq}$ to be close to $1$).

Now the calculation of $\sigma(b)$ shown in (\ref{sigma}) reduces to
straightforward gaussian integrations. The relevant formula is given in the Appendix.

We have analyzed the data at all ISR energies \cite{elastic,p(0)}. It turns
out that the model works very well indeed, thus it is flexible enough. Note
that this is not entirely trivial conclusion. For instance, an analogous
calculation performed in the model with the assumption that the proton
consists of three uncorrelated constituent quarks led to negative conclusions
\cite{pp-3q}.

\begin{figure}[h]
\begin{center}
\includegraphics[scale=1.05]{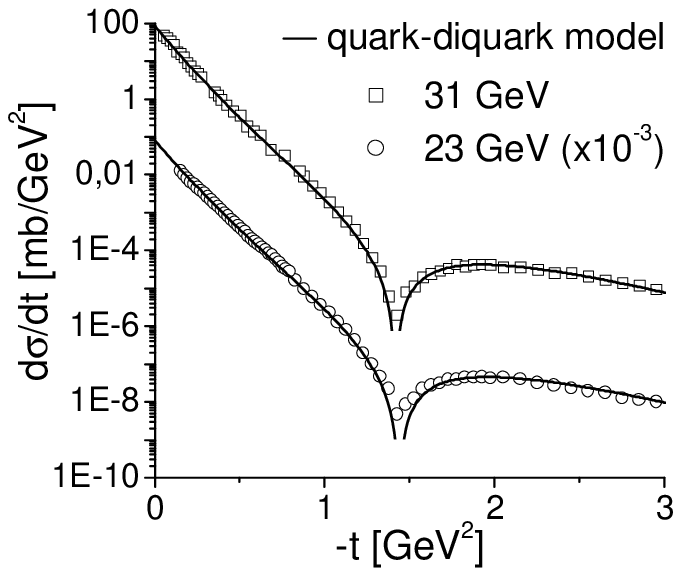}  \hspace{0.2cm}
\includegraphics[scale=1.05]{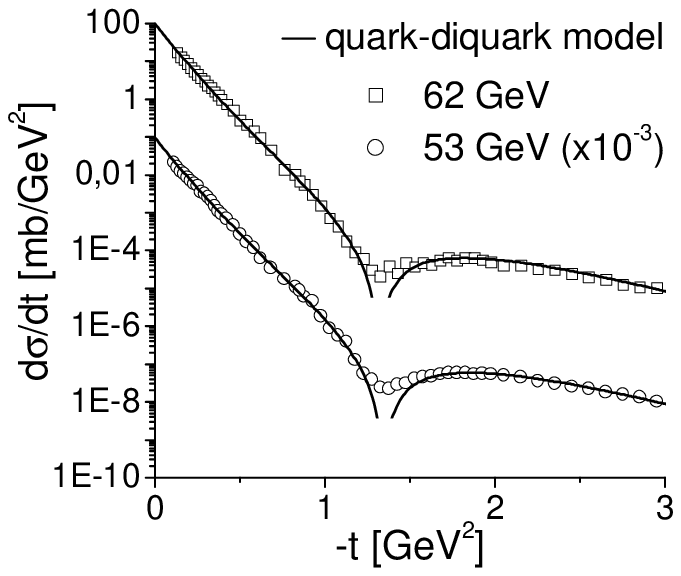}
\end{center}
\caption{The quark-diquark model compared to data on differential cross
section at four energies. Diquark as a simple constituent.}%
\label{Fig_a}%
\end{figure}

Four examples of our calculations are shown in Fig. \ref{Fig_a}, where the
differential cross section $d\sigma/dt$ at the ISR energies $23$, $31$, $53$
and $62$ GeV, evaluated from the model, are compared with data \cite{elastic}%
.

One sees an impressive agreement. Note that we are adjusting the model to
account only for the slope and the value at $t=0$, and for the position of the
minimum. Nevertheless, one sees that the resulting curve follows very well the
subtle structure of the cross-section between $t=0$ and the minimum. This
indicates that the nucleon model with two very different constituents may
indeed represent more than a simple parametrization of data.

The values of the parameters at various energies are given in the Table
\ref{Tab_a}.\footnote{The values of $A_{qq}$ and $\lambda$ are correlated.
Those given in the table are obtained by demanding $A_{qq}=1$.} One sees some
tendency for all radii to increase with increasing energy. Given the
experimental and theoretical inaccuracies the effect is barely
significant.\begin{table}[h]
\begin{center}%
\begin{tabular}
[c]{|c|c|c|c|c|c|}\hline\hline
$\sqrt{s}$ [GeV] & $\lambda$ & $R_{q}$ [fm] & $R_{d}$ [fm] & $R$ [fm] &
$A_{qq}$\\\hline\hline
$23$ & 0.64 & 0.275 & 0.739 & 0.312 & 1\\\hline
$31$ & 0.64 & 0.279 & 0.752 & 0.316 & 1\\\hline
$53$ & 0.71 & 0.288 & 0.770 & 0.327 & 1\\\hline
$62$ & 0.71 & 0.290 & 0.774 & 0.327 & 1\\\hline
\end{tabular}
\end{center}
\caption{Diquark as a simple constituent. The parameters of the model at four
different energies.}%
\label{Tab_a}%
\end{table}

The most striking feature seen in the Table \ref{Tab_a} is the large value of
the diquark radius $R_{d}$. It is almost three times larger than the radius of
the quark, and not very much smaller than the radius of the proton itself. It
is interesting that this feature agrees with other estimates \cite{Wilczek},
based on rather different arguments.

\section{Diquark as a qq system}

In this section we consider another option, treating the diquark as a system
composed of the two constituent quarks. As before we parametrize $\sigma
_{qq}(s)$ using the same simple Gaussian as in (\ref{pab}).
\begin{equation}
\sigma_{qq}(s)=A_{qq}e^{-s^{2}/2R_{q}^{2}}. \label{pqq}%
\end{equation}

To evaluate quark-diquark and diquark-diquark cross-sections we need the
distribution of the quarks inside the diquark which we again take as a
Gaussian:
\begin{equation}
D(s_{q1},s_{q2})=\frac{1}{\pi d^{2}}e^{-\left(  s_{q1}^{2}+s_{q2}^{2}\right)
/2d^{2}}\delta^{2}(s_{q1}+s_{q2}),
\end{equation}
where $s_{q1}$ and $s_{q2}$ are transverse positions of the quarks in the diquark.

Using the Glauber \cite{Glauber} and Czyz-Maximon \cite{cm} expansions,
analogous to (\ref{sigma}), we have:%
\begin{equation}
\sigma_{qd}(s)=\frac{4A_{qq}R_{q}^{2}}{R_{d}^{2}+R_{q}^{2}}e^{-s^{2}\tfrac
{1}{R_{d}^{2}+R_{q}^{2}}}-\frac{A_{qq}^{2}R_{q}^{2}}{R_{d}^{2}}e^{-s^{2}%
/R_{q}^{2}}, \label{pdq}%
\end{equation}%
\begin{align}
\sigma_{dd}(s)  &  =\frac{4A_{qq}R_{q}^{2}}{R_{d}^{2}}e^{-s^{2}\tfrac
{1}{2R_{d}^{2}}}-\frac{4A_{qq}^{2}R_{q}^{4}}{R_{d}^{4}}e^{-s^{2}/R_{d}^{2}%
}-\frac{2A_{qq}^{2}R_{q}^{2}}{2R_{d}^{2}-R_{q}^{2}}e^{-s^{2}/R_{q}^{2}%
}\label{pdd}\\
&  +\frac{4A_{qq}^{3}R_{q}^{4}}{R_{d}^{2}(2R_{d}^{2}-R_{q}^{2})}%
e^{-s^{2}\tfrac{2R_{d}^{2}+R_{q}^{2}}{2R_{q}^{2}R_{d}^{2}}}-\frac{A_{qq}%
^{4}R_{q}^{4}}{(2R_{d}^{2}-R_{q}^{2})^{2}}e^{-s^{2}\tfrac{2}{R_{q}^{2}}%
},\nonumber
\end{align}
where%
\[
R_{d}^{2}=d^{2}+R_{q}^{2}%
\]
is the effective diquark radius.

Introducing this result into the general formulae given in Section 2 one can
evaluate the differential and total inelastic pp cross-sections.

\begin{figure}[h]
\begin{center}
\includegraphics[scale=1.05]{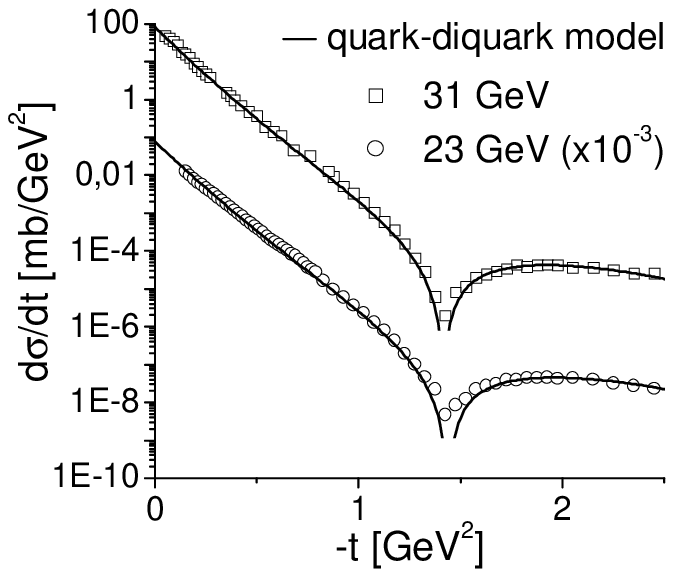}  \hspace{0.2cm}
\includegraphics[scale=1.05]{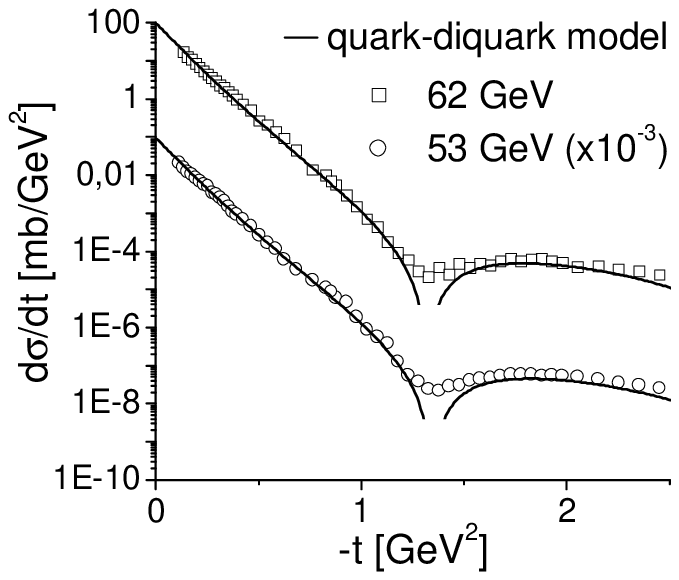}
\end{center}
\caption{The quark-diquark model compared to data on differential cross
section at four energies. Diquark as a $qq$ system.}%
\label{Fig_b}%
\end{figure}

Again we have analyzed the data at all ISR energies \cite{elastic,p(0)}. It
turns out that the model in this form also works reasonably well. The results
are shown in Fig. \ref{Fig_b}, where the differential cross-section
$d\sigma/dt$ at the various ISR energies (the same as in Fig. \ref{Fig_a}) are
compared with the data. There is little difference between plots in Fig.
\ref{Fig_b} and that in Fig. \ref{Fig_a}, except the region $-t>2.5$ GeV$^{2}$
which is of no interest in the present context.

The parameters used in Fig. \ref{Fig_b} are given\footnote{The results are
insensitive to the value of $\lambda$. We have taken $\lambda=1/2$, conforming
to the idea that the binding of quarks inside the diquark is rather weak.} in
Table \ref{Tab_b}.\begin{table}[h]
\begin{center}%
\begin{tabular}
[c]{|c|c|c|c|c|c|}\hline\hline
$\sqrt{s}$ [GeV] & $\lambda$ & $R_{q}$ [fm] & $R_{d}$ [fm] & $R$ [fm] &
$A_{qq}$\\\hline\hline
$23$ & 1/2 & 0.322 & 0.748 & 0.154 & 0.78\\\hline
$31$ & 1/2 & 0.327 & 0.761 & 0.157 & 0.78\\\hline
$53$ & 1/2 & 0.335 & 0.781 & 0.161 & 0.79\\\hline
$62$ & 1/2 & 0.336 & 0.786 & 0.163 & 0.80\\\hline
\end{tabular}
\end{center}
\caption{Diquark as a $qq$ system. The parameters of the model at four
different energies.}%
\label{Tab_b}%
\end{table}

One sees the similar features as in the model of Section 3. The diquark is
much larger than the quark (confirming a weak binding), and the distance
between quark and diquark even smaller than that in previous calculation.

From (\ref{pqq}), (\ref{pdq}) and (\ref{pdd}) one can evaluate the total
inelastic cross-section of the constituents. It turns out that the ratio of
the cross-sections of the constituents satisfies:
\begin{equation}
\sigma_{qq}:\sigma_{qd}:\sigma_{dd}\approx1:1.9:3.7,
\end{equation}
which is rather close to (\ref{124}), indicating a small amount of shadowing.

\section{ Discussion and conclusions}

Our main conclusions can be summarized in three points.

(i) The constituent quark-diquark structure of the nucleon can account very
well for the data on elastic $pp$ scattering at low momentum transfers and c.
m. energies above $20$ GeV.

(ii) The confrontation with data allows to determine the parameters
characterizing the two nucleon constituents.

(iii) The radius of the constituent diquark turns out much larger than that of
the constituent quark. It is comparable to the radius of the nucleon.

Several comments are in order.

(a) It seems remarkable that our calculation reconstructs precisely the fine
structure of the elastic scattering cross-section in the region before the
first minimum. This indicates that, indeed, the two components of the proton
with rather different radii are needed to explain the details of data.

(b) It is reassuring that our conclusion about the large radius of the diquark
agrees with that obtained in \cite{Wilczek} from rather different arguments.

(c) We have verified that the quark-diquark model gives also a good
description of the $\pi p$ scattering. The discussion of this problem will be
subject of a separate investigation \cite{b}.

(d) Finally, we find it significant that the quark-quark and quark-diquark
cross-sections obtained here, when used in the wounded constituent model
\cite{bb}, explain very well the RHIC data on particle production in the
central rapidity region.

Given all these arguments, it seems that the quark-diquark model of the
nucleon structure at low momentum transfers does indeed capture the main
features of this problem and thus deserves a closer attention.

\bigskip

\textbf{Acknowledgements}

We thank R. Peschanski, M. Praszalowicz and G. Ripka for discussions. This
investigation was supported in part by the MEiN Grant No 1 P03 B 04529 (2005-2008).

\section{Appendix}

The presence of the two $\delta$ functions [c.f. (\ref{D})] reduces
(\ref{sigma}) to two gaussian integrations with the substitution
\begin{equation}
s_{d}=-\lambda s_{q};\quad s_{d}^{\prime}=-\lambda s_{q}^{\prime}.
\end{equation}

The integration gives%
\begin{gather}
\frac{4v^{2}}{\pi^{2}}\int d^{2}s_{q}d^{2}s_{q}^{\prime}e^{-2v(s_{q}^{2}%
+s_{q}^{\prime2})}e^{-c_{qq}(b-s_{q}+s_{q}^{\prime})^{2}}e^{-c_{qd}%
(b-s_{q}+s_{d}^{\prime})^{2}}\times\\
\times e^{-c_{dq}(b-s_{d}+s_{q}^{\prime})^{2}}e^{-c_{dd}(b-s_{d}+s_{d}%
^{\prime})^{2}}=\frac{4v^{2}}{\Omega}e^{-b^{2}\Gamma/\Omega},\nonumber
\end{gather}
where:%
\begin{align}
\Omega &  =\left[  4v+\left(  1+\lambda\right)  ^{2}\left(  c_{qd}%
+c_{dq}\right)  \right]  \left[  v+c_{qq}+\lambda^{2}c_{dd}\right]  +\\
&  +\left(  1-\lambda\right)  ^{2}\left[  v\left(  c_{qd}+c_{dq}\right)
+\left(  1+\lambda\right)  ^{2}c_{qd}c_{dq}\right]  ,\nonumber
\end{align}%
\begin{align}
\Gamma &  =\left[  4v+\left(  1+\lambda\right)  ^{2}\left(  c_{qd}%
+c_{dq}\right)  \right]  \left[  v\left(  c_{qq}+c_{dd}\right)  +\left(
1+\lambda\right)  ^{2}c_{qq}c_{dd}\right]  +\\
&  +\left[  4v+\left(  1+\lambda\right)  ^{2}\left(  c_{qq}+c_{dd}\right)
\right]  \left[  v\left(  c_{qd}+c_{dq}\right)  +\left(  1+\lambda\right)
^{2}c_{qd}c_{dq}\right]  .\nonumber
\end{align}

Other integrals can be obtained by putting some of the $c_{ab}=0$.


\begin{thebibliography}{9}                                                                                                %

\bibitem {mm}See, e.g., D. H. Perkins, Introduction to High Energy Physics,
Cambridge U. Press (Cambridge, 2000).

\bibitem {el}See, e.g., H. J. Lipkin, F. Scheck, Phys. Rev. Lett. 16 (1966)
71; E. Levin, L. Frankfurt, JETP Lett. 2 (1965) 65; A. Bialas, K. Zalewski,
Nucl. Phys. B6 (1968) 465.

\bibitem {bcf}A. Bialas, W. Czyz, W. Furmanski, Acta Phys. Pol. B8 (1977) 585;
A. Bialas, W. Czyz, L. Lesniak, Phys. Rev. D25 (1982) 2328.

\bibitem {reszta}R. Nouicer, AIP Conf. Proc. 828 (2006) 11, nucl-ex/0512044;
S. Eremin, S. Voloshin, Phys. Rev. C67 (2003) 064905; P. K. Netrakanti, B.
Mohanty, Phys. Rev. C70 (2004) 027901; Bhaskar De, S. Bhattacharyya, Phys.
Rev. C71 (2005) 024903. See also E. K. G. Sarkisyan, A. S. Sakharov, AIP Conf.
Proc. 828 (2006) 35 and hep-ph/0410324 where the quark model is used to
evaluate the energy deposition in the collision.

\bibitem {bj}See, e.g., J. Bjorken, Acta Phys. Polon. B23 (1992) 637.

\bibitem {pp-3q}A. Bialas, K. Fialkowski, W. Slominski, M. Zielinski, Acta
Phys. Pol. B8 (1977) 855.

\bibitem {Wilczek}See, e.g., R. Jaffe, F. Wilczek, Phys. World 17 (2004) 25;
Phys. Rev. Lett. 91 (2003) 232003; M. Cristoforetti, P. Faccioli, G. Ripka, M.
Traini, Phys. Rev. D71 (2005) 114010.

\bibitem {phobos}B. B. Back et al., Phys. Rev. C65 (2002) 061901; Phys. Rev.
C70 (2004) 021902; Phys. Rev. C72 (2005) 031901; Phys. Rev. C74 (2006) 021901.

\bibitem {bbc}A. Bialas, M. Bleszynski, W. Czyz, Nucl. Phys. B111 (1976) 461.

\bibitem {bb}A. Bialas, A. Bzdak, nucl-th/0611021.

\bibitem {cm}W. Czyz, L. C. Maximon, Ann. of Phys. 52 (1969) 59.

\bibitem {Glauber}R. J. Glauber, Lectures in Theoretical Physics, Vol. 1.
Interscience, New York 1959.

\bibitem {elastic}E. Nagy et al., Nucl. Phys. B150 (1979) 221; N. Amos et al.,
Nucl. Phys. B262 (1985) 689; A. Breakstone et al., Nucl. Phys. B248 (1984)
253; A. Bohm et al., Phys. Lett. B49 (1974) 491.

\bibitem {p(0)}U. Amaldi, K. R. Schubert, Nucl. Phys. B166 (1980) 301.

\bibitem {b}A. Bzdak, to be published.
\end{thebibliography}
\end{document}